\documentclass{vgtc}                          

\ifpdf
  \pdfoutput=1\relax                   
  \usepackage{graphicx}                
  \DeclareGraphicsExtensions{.pdf,.png,.jpg,.jpeg} 
\else
  \usepackage{graphicx}                
  \DeclareGraphicsExtensions{.eps}     
\fi%

\graphicspath{{figures/}{pictures/}{images/}{./}} 

\usepackage{microtype}                 
\PassOptionsToPackage{warn}{textcomp}  
\usepackage{textcomp}                  
\usepackage{mathptmx}                  
\usepackage{times}                     
\usepackage{cite}                      
\usepackage{tabu}                      
\usepackage{booktabs}                  

\usepackage{algorithm}
\usepackage[noend]{algpseudocode}
\usepackage{float}

\makeatletter
\def\algbackskip{\hskip-\ALG@thistlm}
\makeatother

\usepackage{caption}
\usepackage{balance}

\onlineid{0}

\vgtccategory{Research}

\vgtcinsertpkg

\title{Hi-D Maps: An Interactive Visualization Technique\\for Multi-Dimensional Categorical Data}

\author{Radi Muhammad Reza\thanks{e-mail: rreza@ncsu.edu}\\ %
        \scriptsize North Carolina State University %
\and Dr. Benjamin A. Watson\thanks{e-mail: bwatson@ncsu.edu}\\ %
     \scriptsize North Carolina State University}

\begin{document}

\teaser{
    \centering
    \includegraphics[width=7in]{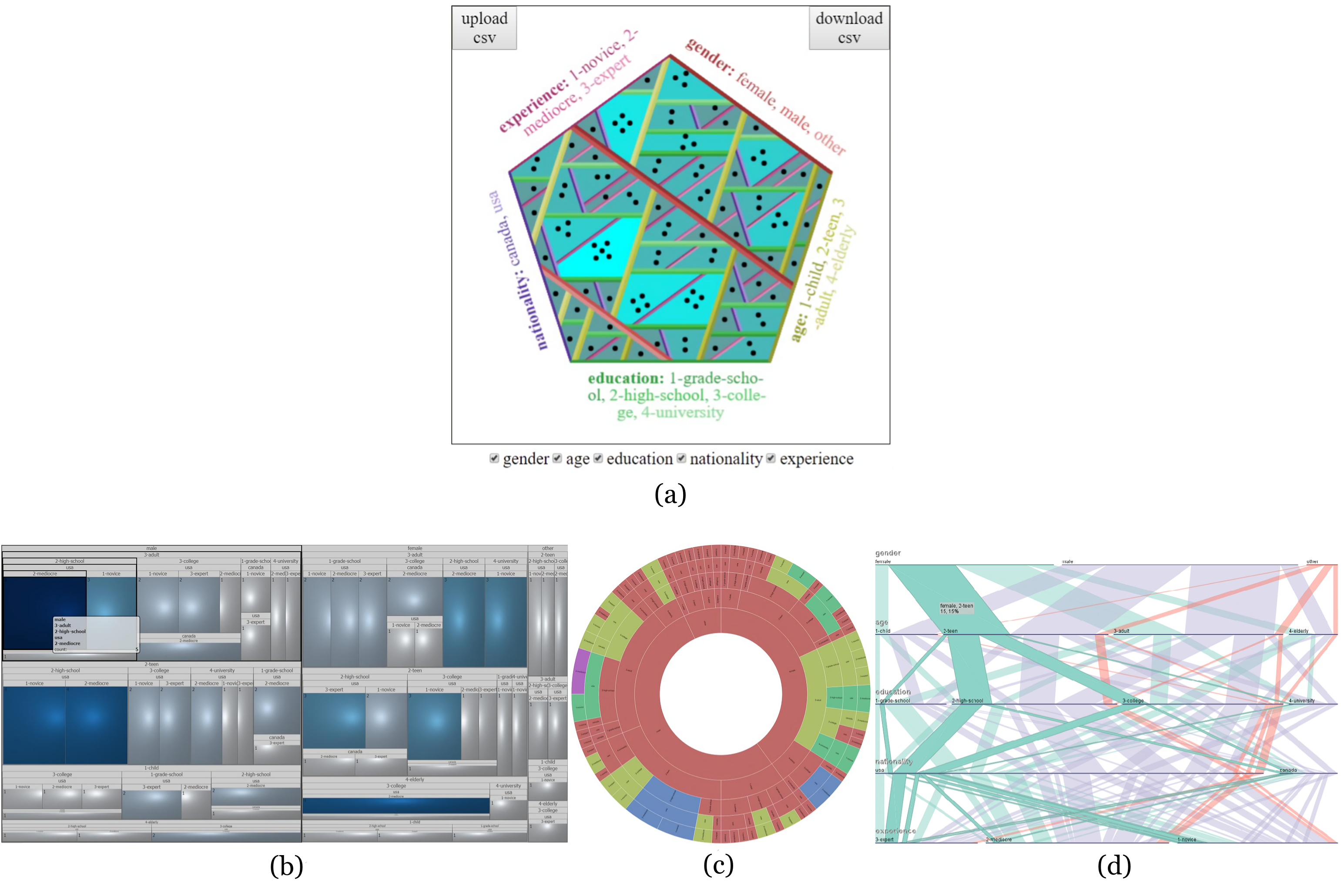}
    \caption{(a) Our \textit{Hi-D maps}, (b) treemaps \cite{johnson1991tree}, (c) sunburst \cite{stasko2000focus+}, and (d) parallel sets \cite{bendix2005parallel} visualizations of the same data.}
   \label{all}
}

\abstract{In this paper, we present \textit{Hi-D maps}, a novel method for the visualization of multi-dimensional categorical data. Our work addresses the scarcity of techniques for visualizing a large number of data-dimensions in an effective and space-efficient manner. We have mapped the full data-space onto a 2D regular polygonal region. The polygon is cut hierarchically with lines parallel to a user-controlled, ordered sequence of sides, each representing a dimension. We have used multiple visual cues such as orientation, thickness, color, countable glyphs, and text to depict cross-dimensional information. We have added interactivity and hierarchical browsing to facilitate flexible exploration of the display: small areas can be scrutinized for details. Thus, our method is also easily extendable to visualize hierarchical information. Our glyph animations add an engaging aesthetic during interaction. Like many visualizations, Hi-D maps become less effective when a large number of dimensions stresses perceptual limits, but Hi-D maps may add clarity before those limits are reached.
} 


\CCScatlist{
  \CCScatTwelve{Human-centered computing}{Visu\-al\-iza\-tion}{Visu\-al\-iza\-tion techniques}{};
  \CCScatTwelve{Human-centered computing}{Visu\-al\-iza\-tion}{Visualization design and evaluation methods}{}
}





\firstsection{Introduction}
\label{intro}

\maketitle

We encounter multi-dimensional categorical data everywhere: in business, social, scientific, and engineering data. Simple, easy to understand tools to visualize such information may help users gain insight, which may in turn facilitate decision-making. To gain insight on categorical data, it is of paramount importance to grasp how the data attributes interact. For example, in demographic data, a typical question might be: ``which age-group of which gender of which ethnicity is the least educated?'' Querying to extract such information from tabular data is time-consuming, and many users will avoid it. Instead, categorical data visualization strives to provide viewers with an overview of the data at a first glance, and to facilitate deeper analyses through focused interaction. Unfortunately, tools for visualization of high-dimensional categorical data are scarce. In this paper, we address this limitation by proposing \textit{Hi-D maps}, a novel tool for visualization of such data.

\section{Related Work}
\label{relatedwork}

We examined several surveys on multi-dimensional visualization techniques \cite{chan2006survey, grinstein2001high, keim1997visual} to find related work, and also sought more recent work. Visualizing multi-dimensional data relies on the basic principle of mapping each dimension to some visual feature in the display. Software packages often extend well-known 1D or 2D visualizations such as histograms, bar charts, and scatter-plots for higher dimensional data by leveraging hue, depth, size, and shape. While being an excellent strategy to depict a few dimensions, such extensions fail for a larger number of dimensions because of perceptual limitations and a lack of available visual features.

Several useful spatial techniques such as table lens \cite{rao1994table}, landscapes \cite{wright1995information}, and Andrews curves \cite{andrews1972plots} depict high-dimensional numeric data \cite{chan2006survey, grinstein2001high, keim1997visual}. They often also vary color, contrast, depth, and shape to illustrate data. However, when used to represent high-dimensional categorical information, spatial mappings lose resolution, and viewers have difficulty making sense of the data. Angular techniques such as radial \cite{hoffman2000table} and star coordinates \cite{kandogan2000star} can help detect clusters in multivariate data, but again have limited utility with categorical information \cite{chan2006survey}. Parallel coordinates \cite{inselberg1990parallel} have similar limitations, but its analogous visualization, parallel sets, depicts categorical data fairly well \cite{bendix2005parallel}. However, even parallel sets become cluttered with a large number of dimensions.

Icon-based techniques such as Chernoff faces \cite{chernoff1973use}, stick figures \cite{grinstein1989exvis}, and shape-coding \cite{beddow1990shape} map each dimensional-attribute to some visual property of a glyph \cite{chan2006survey, keim1997visual}. For example, a Chernoff face might map two dimensions to the 2D position of a face, and the remaining dimensions to facial properties. Such strategies can be used to represent data of reasonably high dimensionality, and their graphical depictions may be processed pre-attentively. However, since the mapping is not natural, the viewer may require practice to decipher the encoding. Moreover, some glyph features are more salient than others, introducing biases in interpretation. Lastly, icon-based techniques can be difficult to scale to data with many items.

In pixel-based visualizations \cite{chan2006survey, grinstein2001high, keim1997visual} such as pixel bar charts \cite{keim2001pixel}, recursive patterns \cite{keim1995recursive}, and circle segments \cite{ankerst1996circle}, each data-attribute is mapped to a colored pixel, with the pixels for a single item usually placed at the same relative position in each dimensional segment. These visualizations are clutter- and overlap-free and can depict a large number of dimensions. However, pixel-based visualizations can be difficult to use when there are many data items.

Map-based visualizations lay out high dimensional data in color or space. The recent data context map \cite{cheng2015data} uses multi-dimensional scaling based on data and attribute similarity matrices. While data context maps are useful for performing visual queries (i.e. finding data items that match certain attributed configurations), they are not ideal for estimating the relative size of categorical intersections. ColorMap$^{\textrm{ND}}$ \cite{cheng2018colormap} maps high-dimensional data to color. It is useful for mapping categorical data to color palettes, but also cannot help in visualizing the cross-dimensional intersections. A few recent techniques such as D-Map \cite{chen2016d} and E-Map \cite{chen2017map} represent non-numeric social media interaction data, but not categorical data.

\subsection*{Hierarchical Visualizations}
\label{hierarchicalvis}

We draw our main inspiration from hierarchical visualizations \cite{chan2006survey, keim1997visual}, including worlds-within-worlds \cite{feiner1990worlds}, treemaps \cite{johnson1991tree} and sunburst visualizations \cite{stasko2000focus+}. These techniques hierarchically partition space into sub-spaces, primarily for representing hierarchical information such as directory structures. Nevertheless, by mapping single dimensions to hierarchy levels, they can be used to represent multi-dimensional categorical data. However, doing so requires giving an order to the dimensions, with the first dimension at the top level, the second at the next, and so on.

Consider the visualizations in Figure~\ref{all}. We have used a synthetic, demographic dataset of 5 dimensions to compare our Hi-D maps with treemaps \cite{johnson1991tree}, sunburst visualizations \cite{stasko2000focus+}, and parallel sets \cite{bendix2005parallel}. Treemaps partition the space horizontally and vertically, alternating in even-odd steps, creating areas proportional to the subset sizes. With only two orientations, these axis-parallel cuts require level-counting to determine the data dimension, making treemaps particularly hard to decipher as the number of dimensions grows. 

Sunburst visualizations are similar to treemaps, but partition the space radially: the innermost ring represents the first dimension, the outermost the last. Angular extent represents subset size. Sunburst visualizations are less efficient with space than treemaps, since each dimension occupies a different ring. Viewers can easily find items in leaf nodes, but to find a specific intersection of attributes, they must still traverse the hierarchy. 

We have designed Hi-D maps to retain the spatial efficiency of treemaps and dimensional clarity of sunburst visualizations.

\begin{figure}[t]
 \centering
 \includegraphics[width=3.3in]{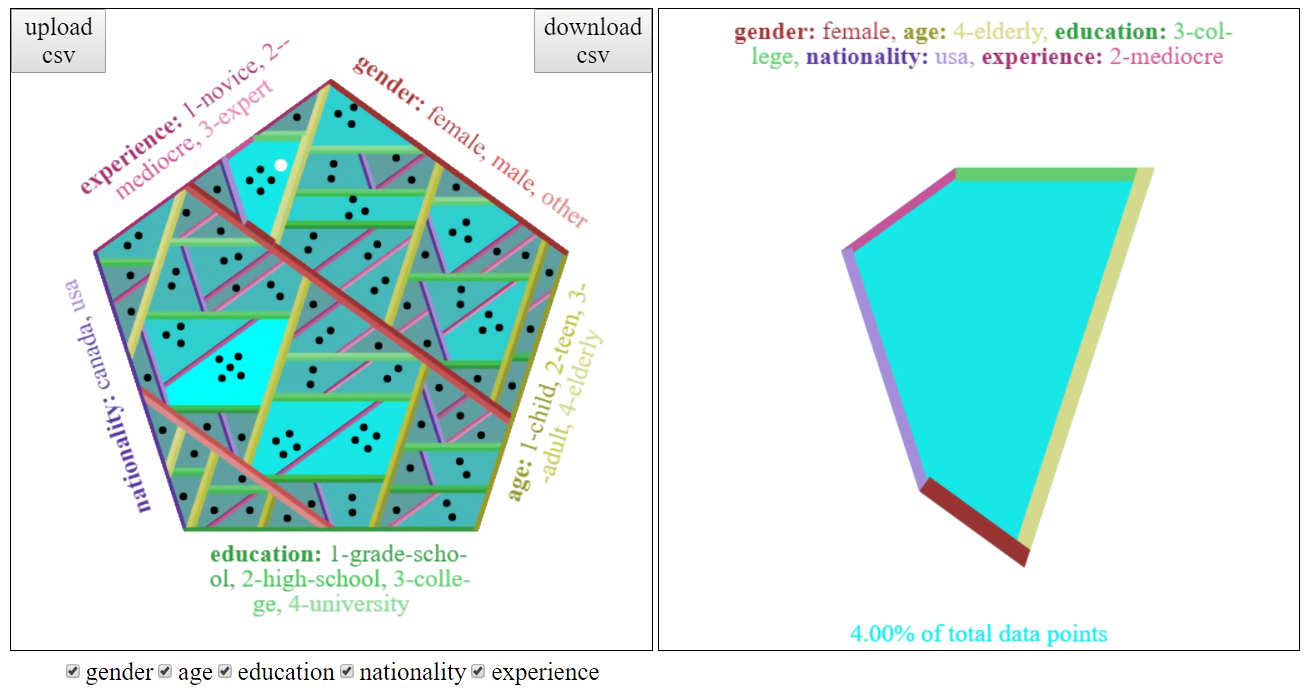}
 \caption{Hi-D Maps: multiple dimensions mapped to a regular polygon. The polygon is cut hierarchically with lines parallel to the sides corresponding to the dimensions. The area of a leaf polygon is proportional to the size of the corresponding data intersection. The full dataset is displayed in the left overview window; as the mouse hovers over a subset, it is magnified in the right detail window.}
 \label{overview}
\end{figure}

\section{Design}
\label{design}

To retain spatial efficiency while also providing a clear dimensional mapping, Hi-D maps apply an alternate space-partitioning strategy, mapping dimensions to the oriented sides of a regular polygon, ordered clockwise from top-right. In Figure~\ref{overview} - left, Hi-D maps first partitions the polygon based on gender using lines parallel to the top-right side, with the region closest to that side depicting the lexicographically smallest categorical value: female. The proportional area of each resulting sub-polygon represents the proportion of all items that belong to that data partition. Each sub-polygon and data partition from this stage is then recursively partitioned in the same manner, based next on age, followed by education, nationality, and experience. Each resulting leaf polygon depicts a five-dimensional intersection with a proportional area matching the proportion of data points in the intersection. Note that when the number of dimensions is even, Hi-D maps avoid parallel sides by keeping the number of sides odd, with no dimension mapped to the last side. Our space utilization is efficient, since an arbitrarily large information structure is mapped to a 2D polygon of constant circumradius.

Hi-D maps provide multiple, often reinforcing, visual cues to reduce the viewer's perceptual and cognitive loads. Hi-D maps reinforce the angular display of dimension with hue in the HSL color space, varying linearly from 0.0 to 0.4, and from 0.6 to 0.9. (We reserve 0.5 to depict area, and avoid 1.0 due to the circular nature of hue). To indicate value ordering in each dimension, we vary lightness from 0.4 to 0.7 with lexicographically smaller values being darker. The text depicting the corresponding value assumes the same lightness. To reinforce display of the proportional amount of data in an intersection, we supplement area with two cues: first, color saturation, which we vary from 0.25 for the smallest to 1.0 for the largest, for quick comparisons; and second, countable glyphs or ``marbles'' inside each leaf polygon with their count equalling the rounded percentage of the total Hi-D maps area occupied by the polygon. Within each polygon, we place the marbles randomly in groups of five (to facilitate counting) with the last group having possibly fewer. Note that the marbles do not represent actual data items; they only help viewers understand the proportional size of each dimensional intersection. Line thickness indicates dimension order, with the first dimension thickest.

\begin{figure}[t]
 \centering
 \includegraphics[width=3.3in]{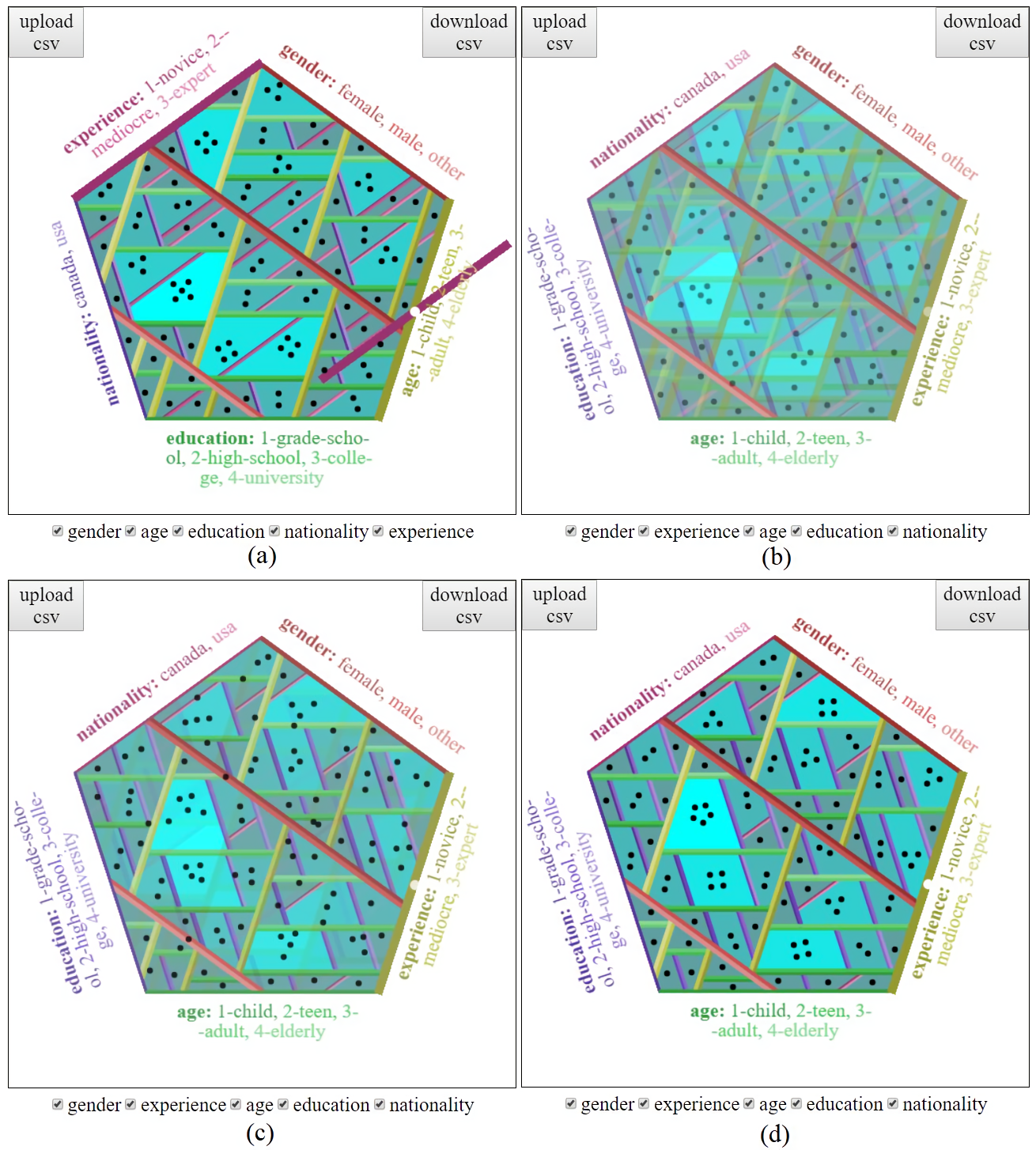}
 \caption{(a) The dimensions can be re-ordered by dragging one edge atop another. (b-c) Hi-D maps animate this change, including movements of the marbles. (d) New display with re-arranged dimensions allowing different interpretations.}
 \label{reorder}
\end{figure}

\subsection{Interaction}

\begin{figure}[tb]
 \centering
 \includegraphics[width=3.3in]{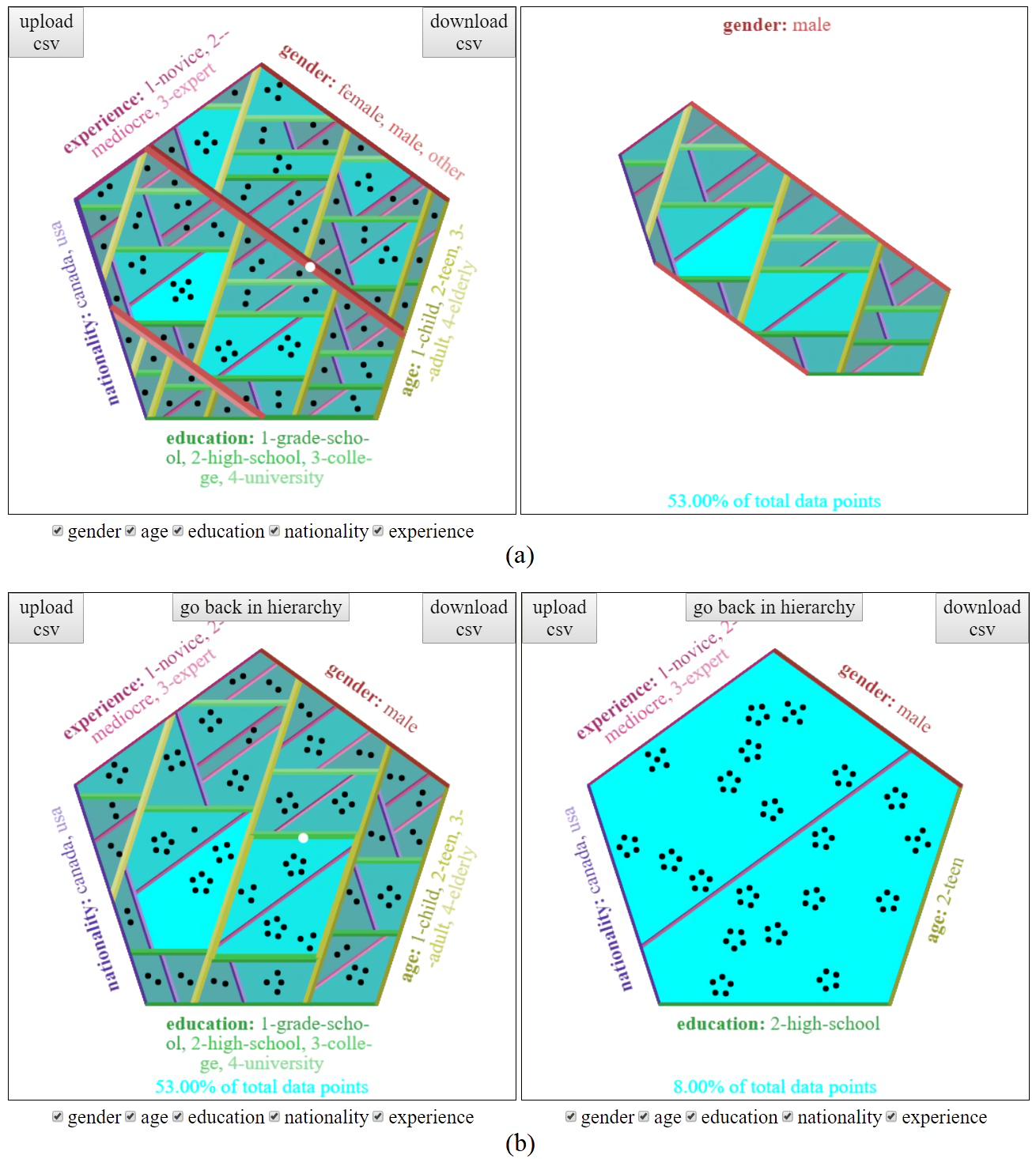}
 \caption{(a) Full dataset and the details of a subset. (b) Hierarchical browsing: first for the subset chosen in (a), then for the next selected subset. The button at the top-middle supports going back.}
 \label{hierarchy}
\end{figure}

When the number of dimensions and values per dimension increase, Hi-D maps may become cluttered. To help viewers when this occurs, we have applied an ``overview and detail'' strategy. The full visualization is in an overview window, with details in a separate window (see Figure~\ref{overview}). To choose the detailed area, the viewer's mouse hovers over a leaf polygon. In addition to the magnified polygon, Hi-D maps shows the percentage of data points in the polygon at the bottom, and the associated dimension-values at the top. Since inner polygons may have fewer boundary-edges than the total number of dimensions, we have decided to print the dimension-values at the top instead of along the edges. When the viewer's mouse hovers over an edge, the detail window displays similar information for the corresponding, possibly non-leaf, node (see Figure~\ref{hierarchy}(a)).

Most multi-dimensional techniques stress uniform-treatment of dimensions because differences such as ordering may alter perceived meaning \cite{chan2006survey}. Since our dimensions are ordered, Hi-D maps allows viewers to re-order them. Viewers may drag one side atop another (see Figure~\ref{reorder}(a)), taking its place in the order, and shifting all the intermediate dimensions by one side (see Figure~\ref{reorder}(d)). This reordering mitigates the perceptual bias introduced by the initial order of the dimensions. Note that the number and sizes of the leaf sub-polygons will remain the same regardless of the chosen ordering.

Inspired by hierarchy visualizations \cite{johnson1991tree, stasko2000focus+}, Hi-D maps support hierarchical filtering and browsing to manage complexity. In Figure~\ref{hierarchy}(a), when the viewer's mouse hovers over the red edge, the detail window shows a subset with unique gender: only male. To zoom in on that portion of the data, the viewer need only click on the edge, filtering out all data with gender not male (see Figure~\ref{hierarchy}(b) - left). For example, Figure~\ref{hierarchy}(b) - right shows the visualization for gender: male, age: teen, and education: high-school.  To go back in hierarchy, the viewer need only press the top button. With this functionality, Hi-D maps might be extended to depict truly hierarchical information, with the innermost polygons representing leaves.

\begin{figure}[tb]
 \centering
 \includegraphics[width=3.3in]{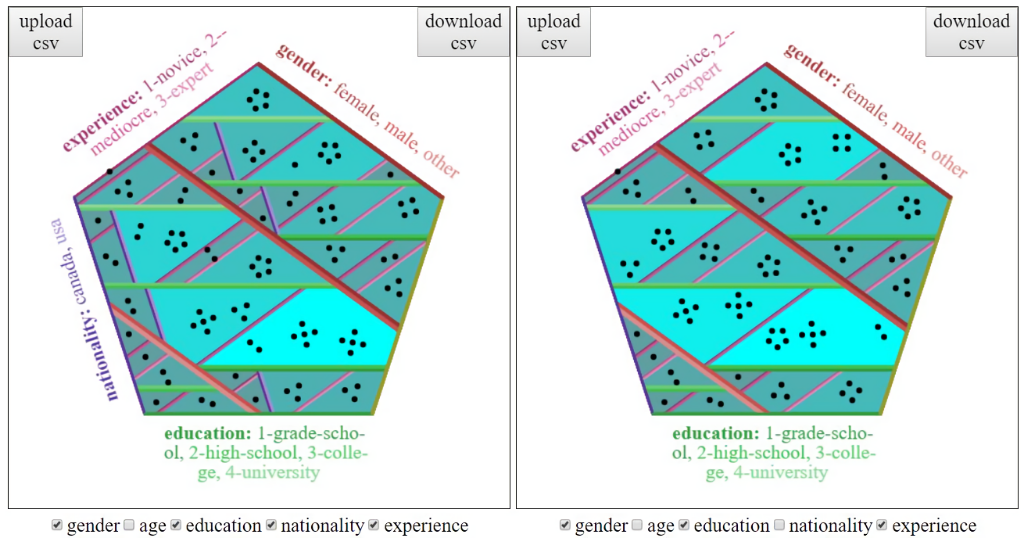}
 \caption{The check-boxes at the bottom supports partial display. From left to right: hiding one and two dimensions respectively.}
 \label{hide}
\end{figure}

Viewers may also manage complexity by hiding uninteresting dimensions, as shown in Figure~\ref{hide}. Viewers may check/uncheck the boxes at the bottom to view/hide each dimension. The other dimensions stay in place; the outer polygon remains the same but the inner polygons change slightly as our algorithm recomputes areas. By keeping the overall Hi-D maps shape, we reduce cognitive load.

Finally, to improve visual continuity and engage the viewer, we have animated all display updates. For example, in Figure~\ref{reorder}(b-c), we show two transitional animation frames during a dimension re-ordering. Notice that the previous display fades out as the new one fades in. Some marbles change position through smooth non-overlapping physical movements. To simulate marble-movements, we first calculate their new placements. Then we greedily assign each marble the task of moving towards the nearest available new position. We could use an expensive min-cost flow routine to minimize the total movement of the marbles, however, our greedy strategy works nearly as well. To re-position the marbles smoothly, we have used an agent-based simulation technique \cite{benson2008agentbased} which updates the velocity of each marble per frame by computing a current distance matrix and an ideal distance matrix between the marbles and their destinations. To avoid overlaps, we have added extra rules. As a result, the marbles move fluidly, avoiding when they approach each other. Our refreshing animations may alleviate boredom and engage viewers as they interactively browse Hi-D maps.

\subsection{Comparison}

We believe Hi-D maps have many of the strengths of other hierarchical visualizations, and few of the weaknesses. They are space-efficient like treemaps (Figure~\ref{all}(b)), but do not suffer as much from dimensional ambiguity, since they display dimension with angle and hue. Like sunburst visualizations (Figure~\ref{all}(c)), dimensional intersections are easy to find, but especially so because all dimensions share the same space. Unlike parallel sets (Figure~\ref{all}(d)), finding cross-dimensional intersections does not require path-following through complex, intersecting links. 

\section{Algorithm}
\label{algo}

In this section, we present our main space-partitioning algorithm. The augmenting features we illustrated in Section 3 need only basic tracking and either partial or complete re-computation of the visualization; we omit these algorithms due to space constraints.

\begin{algorithm}
\caption{Hi-D Maps}\label{algo}
\begin{algorithmic}[1]
    \Procedure{Main}{$display,\ mouse$}
        \State $data \gets$ load from file
        \State $rootPoly \gets$ \Call{Init}{$data.dim,\ data.attrib$}
        \State $tree \gets$ \Call{BuildTree}{$rootPoly,\ data,\ 0$}
        \While{true}
            \State \Call{Update}{$display,\ mouse,\ tree$}
            \State \Call{Draw}{$display,\ tree$}
        \EndWhile
    \EndProcedure
    \Procedure{BuildTree}{$poly,\ data,\ dimension$}
        \State $treeNode \gets$ \Call{MakeNode}{$poly,\ data.attrib,\ dimension$}
        \If{$dimension = data.dim.length$}
            \Return $treeNode$
        \EndIf
        \State $partitionedData \gets$ \Call{Partition}{$data,\ dimension$}
        \State $splitPolys \gets$ \Call{SplitByBinarySearch}{}($poly,\ dimension.$\par\hspace{2cm}$slope,\ partitionedData$)
        \For{$i \gets$ 0 \ \textbf{to} \ $partitionedData.length - 1$}
            \State $child \gets$ \Call{BuildTree}{}($splitPolys[i],\ partitionedData[i],$\par\hspace{1.8cm}$dimension + 1$)
            \State $treeNode.$\Call{AddChild}{$child$}
        \EndFor
        \State \Return $treeNode$
    \EndProcedure
\end{algorithmic}
\end{algorithm}

\newpage

Algorithm~\ref{algo} details our approach. The \textsc{Main} procedure loads the data from a file in line 2. Line 3 initializes the bounding regular polygon and associated text labels. Line 4 calls the \textsc{BuildTree} procedure which recursively computes the tree, representing Hi-D maps. At its first call, \textsc{BuildTree} initializes the root node with the bounding regular polygon in line 9. Line 11 partitions the current data based on the values in the passed dimension, which initially is the first one. In line 12, the polygon is split into smaller polygons, each of area proportional to one of the data-partitions. The for-loop in lines 13-15 recursively computes children nodes for the next dimension and the partitioned data and polygons from earlier. Line 16 returns the node. The recursion stops when all dimensions are used. In the \textsc{Main} procedure, an animation loop in lines 5-7 constantly waits for feature-updates through mouse events and redraws the display upon necessity. 

Ignoring the animation loop, the data-partitioning and the binary search to split polygons in the recursive step dominates the time complexity of the algorithm. Let us assume that the data has $m$ rows and $n$ dimensions, each dimension has maximum $p$ values, and any edge has maximum $l$ length. Thus, the time complexity of our algorithm is $O(n(m + p) + p^nnlgl)$. Here, rather than the size of the dataset $m$, the exponential $p^n$ term dominates the complexity and determines the computational scalability. Thus, our visualization is suitable for categorical data with a small number of values per dimension, not for numerical data.

\section{Conclusions and Future Work}
\label{concl}

We have presented a novel interactive visualization technique called Hi-D maps to represent multi-dimensional categorical data. Our methodology is space-efficient. We have provided several augmenting features supporting detail-view, dimension re-ordering, hierarchical browsing, and dimension-hiding. These along with interaction and fluid animation make our visualization helpful in analyzing data of interest quickly, effectively, and in an engaging manner.

Despite all its advantages, our design suffers from a number of limitations. When the number of dimensions becomes large (in our experience, at about 30), the polygon will closely approximate a circle, dimensional orientations and hues will be difficult to distinguish, and our technique will fail. Large numbers of values in each dimension magnifies this problem, creating very small inner leaf polygons. In such polygons, the proportional size of intersections will be difficult to perceive, with saturations quite similar, and marbles no longer fitting (we use 100 marbles, meaning each polygon must have at least 0.5\% of data items). Amid such clutter, polygon boundaries and text may also no longer fit.

In future work, we may conduct a series of user studies evaluating the understandability, aesthetics, usability and insightfulness of Hi-D maps in both synthetic and real application scenarios. We plan to extend our technique to visualize hierarchical information as well. Later, we may empirically compare our technique with other tools. We may also address several of the limitations we highlighted above. To resolve clutter, we may consider a pan and zoom strategy in the overview window. When zooming an area, we may recompute thickness and color values for that area to distinguish nearby dimensions and attributes. Instead of printing all values of a dimension, we may list them as a drop-down menu. We may allow the viewer to change the ordering of the values in any dimension by re-ordering the items in the drop-down menu. Of course, like any design, Hi-D maps reflect a trade-off amid perceptual (e.g., of angle, size and color) and data (categorical, nominal/ordinal/ratio-interval) constraints; we cannot eliminate these limitations altogether.

\acknowledgments{
The authors wish to thank Brighid Wilhite, Yuhao Xu, Sixing Hu and Rui Bai, who helped develop and evaluate earlier versions of Hi-D maps.}

\balance

\bibliographystyle{abbrv-doi}

\bibliography{template}

\begin{thebibliography}{10}

\bibitem{andrews1972plots}
D.~F. Andrews.
\newblock Plots of high-dimensional data.
\newblock {\em Biometrics}, pp. 125--136, 1972.

\bibitem{ankerst1996circle}
M.~Ankerst, D.~A. Keim, and H.-P. Kriegel.
\newblock Circle segments: A technique for visually exploring large
  multidimensional data sets.
\newblock In {\em Visualization}, 1996.

\bibitem{beddow1990shape}
J.~Beddow.
\newblock Shape coding of multidimensional data on a microcomputer display.
\newblock In {\em Proceedings of the 1st Conference on Visualization'90}, pp.
  238--246. IEEE Computer Society Press, 1990.

\bibitem{bendix2005parallel}
F.~Bendix, R.~Kosara, and H.~Hauser.
\newblock Parallel sets: visual analysis of categorical data.
\newblock In {\em IEEE Symposium on Information Visualization, 2005. INFOVIS
  2005.}, pp. 133--140. IEEE, 2005.

\bibitem{benson2008agentbased}
J.~R. Benson, D.~Crist, P.~Lafleur, and B.~Watson.
\newblock Agentbased visualization of streaming text.
\newblock In {\em I7 Information Visualization Conference}, 2008.

\bibitem{chan2006survey}
W.~W.-Y. Chan.
\newblock A survey on multivariate data visualization.
\newblock {\em Department of Computer Science and Engineering. Hong Kong
  University of Science and Technology}, 8(6):1--29, 2006.

\bibitem{chen2017map}
S.~Chen, S.~Chen, L.~Lin, X.~Yuan, J.~Liang, and X.~Zhang.
\newblock E-map: A visual analytics approach for exploring significant event
  evolutions in social media.
\newblock In {\em 2017 IEEE Conference on Visual Analytics Science and
  Technology (VAST)}, pp. 36--47. IEEE, 2017.

\bibitem{chen2016d}
S.~Chen, S.~Chen, Z.~Wang, J.~Liang, X.~Yuan, N.~Cao, and Y.~Wu.
\newblock D-map: Visual analysis of ego-centric information diffusion patterns
  in social media.
\newblock In {\em 2016 IEEE Conference on Visual Analytics Science and
  Technology (VAST)}, pp. 41--50. IEEE, 2016.

\bibitem{cheng2015data}
S.~Cheng and K.~Mueller.
\newblock The data context map: Fusing data and attributes into a unified
  display.
\newblock {\em IEEE transactions on visualization and computer graphics},
  22(1):121--130, 2015.

\bibitem{cheng2018colormap}
S.~Cheng, W.~Xu, and K.~Mueller.
\newblock Colormap nd: A data-driven approach and tool for mapping multivariate
  data to color.
\newblock {\em IEEE transactions on visualization and computer graphics},
  25(2):1361--1377, 2018.

\bibitem{chernoff1973use}
H.~Chernoff.
\newblock The use of faces to represent points in k-dimensional space
  graphically.
\newblock {\em Journal of the American Statistical Association},
  68(342):361--368, 1973.

\bibitem{feiner1990worlds}
S.~Feiner and C.~Beshers.
\newblock Worlds within worlds: metaphors for exploring n-dimensional virtual
  worlds.
\newblock In {\em UIST}, vol.~90, pp. 76--83, 1990.

\bibitem{grinstein1989exvis}
G.~Grinstein, R.~Pickett, and M.~G. Williams.
\newblock Exvis: An exploratory visualization environment.
\newblock In {\em Proc. Graphics Interface}, vol.~89, pp. 254--261, 1989.

\bibitem{grinstein2001high}
G.~Grinstein, M.~Trutschl, and U.~Cvek.
\newblock High-dimensional visualizations.
\newblock In {\em Proceedings of the Visual Data Mining Workshop, KDD}, vol.~2,
  p. 120. Citeseer, 2001.

\bibitem{hoffman2000table}
P.~E. Hoffman.
\newblock {\em Table visualizations: a formal model and its applications}.
\newblock University of Massachusetts Lowell, 2000.

\bibitem{inselberg1990parallel}
A.~Inselberg and B.~Dimsdale.
\newblock Parallel coordinates: a tool for visualizing multi-dimensional
  geometry.
\newblock In {\em Proceedings of the 1st conference on Visualization'90}, pp.
  361--378. IEEE Computer Society Press, 1990.

\bibitem{johnson1991tree}
B.~Johnson and B.~Shneiderman.
\newblock {\em Tree-maps: A space-filling approach to the visualization of
  hierarchical information structures}.
\newblock IEEE, 1991.

\bibitem{kandogan2000star}
E.~Kandogan.
\newblock Star coordinates: A multi-dimensional visualization technique with
  uniform treatment of dimensions.
\newblock In {\em Proceedings of the IEEE Information Visualization Symposium},
  vol. 650, p.~22. Citeseer, 2000.

\bibitem{keim1997visual}
D.~A. Keim.
\newblock Visual techniques for exploring databases.
\newblock In {\em Knowledge Discovery in Databases (KDD'97)}, 1997.

\bibitem{keim1995recursive}
D.~A. Keim, M.~Ankerst, and H.-P. Kriegel.
\newblock Recursive pattern: A technique for visualizing very large amounts of
  data.
\newblock In {\em Proceedings of the 6th Conference on Visualization'95}, p.
  279. IEEE Computer Society, 1995.

\bibitem{keim2001pixel}
D.~A. Keim, M.~C. Hao, J.~Ladisch, M.~Hsu, and U.~Dayal.
\newblock Pixel bar charts: A new technique for visualizing large
  multi-attribute data sets without aggregation.
\newblock In {\em IEEE Symposium on Information Visualization: INFOVIS 2001},
  pp. 113--120, 2001.

\bibitem{rao1994table}
R.~Rao and S.~K. Card.
\newblock The table lens: merging graphical and symbolic representations in an
  interactive focus+ context visualization for tabular information.
\newblock In {\em Proceedings of the SIGCHI conference on Human factors in
  computing systems}, pp. 318--322. Citeseer, 1994.

\bibitem{stasko2000focus+}
J.~Stasko and E.~Zhang.
\newblock Focus+ context display and navigation techniques for enhancing
  radial, space-filling hierarchy visualizations.
\newblock In {\em IEEE Symposium on Information Visualization 2000. INFOVIS
  2000. Proceedings}, pp. 57--65. IEEE, 2000.

\bibitem{wright1995information}
W.~Wright.
\newblock Information animation applications in the capital markets.
\newblock In {\em Proceedings of Visualization 1995 Conference}, pp. 19--25.
  IEEE, 1995.

\end{thebibliography}
\end{document}